# Quantification of sulfated polysaccharides in mouse and rat plasma by the Heparin Red mix-and-read fluorescence assay

Ulrich Warttinger[1], Christina Giese[1], Roland Krämer[1]

Correspondence to:
Roland Krämer, phone 0049 6221 548438, fax 0049 6221 548599
E-mail: kraemer@aci.uni-heidelberg.de

[1] Heidelberg University, Inorganic Chemistry Institute, Im Neuenheimer Feld 270, 60129 Heidelberg, Germany

## Abstract

Sulfated polysaccharides constitute a large and complex group of macromolecules which possess a wide range of important biological properties. Many of them hold promise as new therapeutics, but determination of their blood levels during pharmacokinetic studies can be challenging. Heparin Red, a commercial mix-and-read fluorescence assay, has recently emerged as a tool in clinical drug development and pharmacokinetic analysis for the quantification of sulfated polysaccharides in human plasma. The present study describes the application of Heparin Red to the detection of heparin, a highly sulfated polysaccharide, and fucoidan, a less sulfated polysaccharide, in spiked mouse and rat plasmas. While the standard assay protocol for human plasma matrix gave less satisfactory results, a modified protocol was developed that provides within a detection range 0-10 µg per mL better limits of quantification, 1.1 – 2.3 µg per mL for heparin, and 1.7 – 3.4 µg per mL for fucoidan. The required plasma sample volume of only 20 µL is advantegous in particular when blood samples need to be collected from mice. Our results suggest that Heparin Red is a promising tool for the preclinical evaluation of sulfated polysaccharides with varying sulfation degrees in mouse and rat models.

## Keywords

Sulfated polysaccharides, heparin, fucoidan, Heparin Red, assay, mouse, rat, plasma

## Introduction

Sulfated polysaccharides constitute a large and complex group of macromolecules known to possess a wide range of important biological properties. Heparin (scheme 3, left), a naturally occuring, polydisperse polysaccharide belonging to the glycosaminoglycan family, is of tremendous clinical importance as an anticoagulant drug. It has long been known to have biological effects that are unrelated to its anticoagulant activity, and there has been a recent burgeoning of interest in therapeutic applications of heparins and related sulfated polysaccharides beyond anticoagulant activity.[1,2]

Examples of "non-anticoagulant" sulfated polysaccharides in advanced clinical trials include tafoxiparin [3] (a chemically modified heparin, used for the treatment of prolonged labour), sevuparin [4] (a chemically modified heparin, treatment for sickle cell disease), muparfostat [5] (synthetic sulfated pentasaccharide, liver cancer), pixatimod [6] (highly sulfated synthetic tetrasaccharide, solid tumors), roneparstat [7] (chemically modified heparin, advanced multiple myeloma) and ibsolvmir [8] (dextran sulfate, preventing rejection of transplanted pancreatic islet cells in diabetes patients). Many more sulfated polysaccharides that hold promise as new therapeutics are in earlier or later stages of pre-clinical development, as highlighted by recent reviews.[9-12] For example, algae-derived polysaccharides such as fucoidans (scheme 3, right) have an attractive array of biological activities with potential health benefits.[13-14]

For both clinical and preclinical drug development, pharmacokinetic analysis needs to be considered. A fundamental requirement for understanding the pharmacokinetics of sulfated polysaccharides is the availability of robust analytical methods, in particular for determining the blood levels of the target compound. The structural complexity makes the quantification of sulfated polysaccharides in complex matrices such as blood plasma challenging, in particular if the compounds lack siginificant anticoagulant activity so that the clinically established, indirect heparin assays based on interaction with coagulation factors are not sufficiently sensitive. Methods based on isolation of the sulfated polysaccharides or – if available - enzyme linked immunosorbent assays involve tedious, time consuming multistep protocols and are not first choice for the analysis of large sample numbers in preclinical or clinical trials.

The drawbacks of analytical methods for sulfated polysaccharides have stimulated the development of simple direct detection methods with cationic dyes that change absorbance or fluorescence upon binding of the polyanionic target. [15] Very few such dye-based assays, however, are commercially and widely available to a broad community of researchers and clinicians. Heparin Red is a commercial fluorescent molecular probe for the detection of sulfated polysaccharides in plasma with outstanding sensitivity in the low µg/mL range.[16] It is a polyamine functionalized, red-fluorescent perylene diimide derivative (scheme 1) that strongly binds polyanionic polysaccharides so that aggregation of dye molecules results in contact quenching of fluorescence (scheme 2).[17, 18]

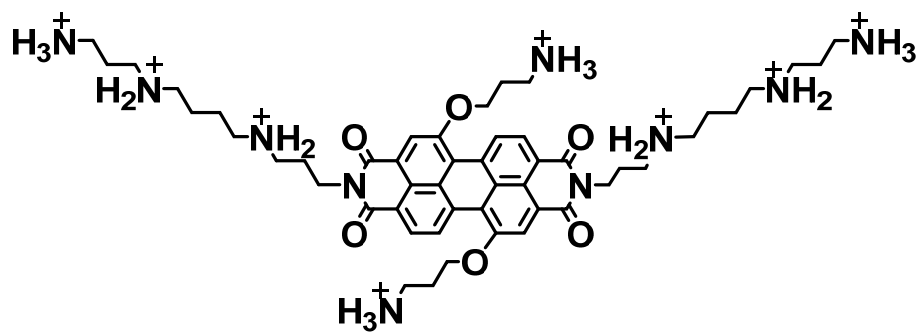

**Scheme 1**. Structure of Heparin Red.

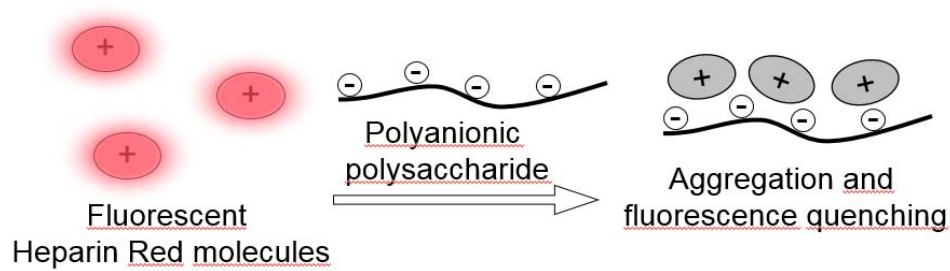

**Scheme 2**. Schematic representation of fluorescence quenching of the molecular probe Heparin Red in the presence of polyanionic polysaccharides.

The commercial Heparin Red Kit has been applied to the sensitive quantification in human plasma of unfractionated and low-molecular-weight heparins [19], heparin octa- and decasaccharides [19], chemically modified heparins including tafoxiparin [19], heparan sulfate [20], algae-derived fucoidans [21], carrageenan [22] and ulvan [22], and the semi-synthetic polysaccharides dextran sulfate [22] and sulfated hyaluronic acid [22]. Heparin Red is used for pharmacokinetic studies in several of the abovementioned clinical trials of sulfated polysaccharide drug candidates.[23] All these applications of Heparin Red have focused so far on the human plasma matrix. For preclinical studies, mice and rats are the most widely used model organisms. Human, mouse and rat plasma differ in protein composition.[24] Human serum albumin, the most abundant plasma protein, shares only about 70% amino acid sequence homology with mouse and rat albumin.[25] Consequently, binding of small molecules and drugs to albumin and other plasma proteins can differ significantly between human and animal species.[26] Another particularity is the small body size and limited blood volume of rats and especially mice. Quantitative assays for drug exposure that require large blood or plasma sample volumes may not comply with animal welfare guidelines.[27]

The present study describes for the first time the application the Heparin Red to the detection of sulfated polysaccharides in mouse and rat plasma. Since response of Heparin Red depends on charge density (sufation degree, respectively) and the polysaccharides species in preclinical development cover a range of charge densities, we have selected as analytical targets heparin, a highly sulfated polysaccharide having an averaged charge of about -1.8 per monosaccharide (scheme 3), and a fucoidan with -0.8 per monosaccharide, representing the less sulfated polysaccharides.

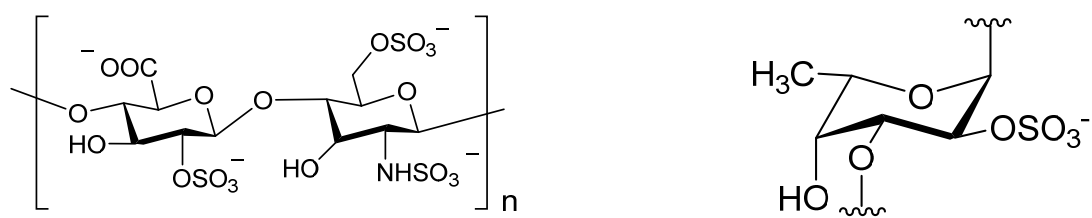

**Scheme 3**. Left: Structure of the major repeating disaccharide unit of heparin. The averaged charge density per monosaccharide is typically -1.8.[28, 29]  Right: Repeating sulfated fucose unit, as present in many fucoidans. Sulfation pattern is variable, fucose may be monosulfated, disulfated or non-sulfated. The averaged charge density per monosaccharide of the fucoidan used in this study (from the brown algae species *F. vesiculosus*) is typically -0.8. [14, 21]

## Materials and Methods

### Instrumentation

*Fluorescence measurements*
Fluorescence (Heparin Red® assay) was measured with a microplate reader Biotek Synergy Mx (Biotek  Instruments, Winooski, VT, USA), excitation at 570 nm, emission recorded at 605 nm,  spectral band width 13.5 nm, read height of 8 mm. Sensitivity of the instrument was adjusted by setting the gain 100 for heparin detection (figure 1-4) and 120 for fucoidan detection (figure 5-8).

*Microplates*
For fluorescence measurements (Heparin Red® assay) 96 well microplates, polystyrene, Item No 655076, were purchased from Greiner Bio-One GmbH, Frickenhausen.

*Pipettes*

Transferpette® 0,5-10µl, Transferpette®-8 20-200µl and Transferpette®-12 20-200µl, purchased from Brand GmbH, Wertheim. Rainin Pipettes 100-1000µl, 20-200µl, and 2-20µl purchased from Mettler Toledo, OH, USA.

**Reagents**

*Heparin Red Kit*

The Heparin Red® Kit was a gift from Redprobes UG, Münster, Germany [30]. Kit components: Heparin Red solution, Product No HR001, Lot 01-003, and Enhancer Solution.

*Sulfated polysaccharides*

Unfractionated heparin sodium salt from porcine intestine mucosa ("heparin"), was sourced as a solid from Sigma-Aldrich GmbH, Steinheim (product number H5515, Lot SLBK0235V, indicated potency 210 IU/mg). Fucoidan from from *Fucus vesiculosus*, purity >95%, product number F8190, Lot SLBN8754V, was purchased from Sigma-Aldrich GmbH, Steinheim.

*Plasma*

Pooled human plasma was prepared by mixing equal volumes of ten single-donor citrated plasmas of healthy individuals, provided by the Blood Bank of the Institute for Clinical Transfusion Medicine and Cell Therapy Heidelberg (IKTZ). Citrated mouse plasma, product number IGMS-N-N02-50, Lot Nr. 22430 (Innovative Grade, US origin) was obtained from Dunn Labortechnik GmbH, Asbach. This plasma was recovered from whole blood donations from normal healthy CD-1 mice. Citrated rat plasma "rat 1", product number RTPC07, Lot Nr. 171023-0207 was obtained from Gentaur Molecular Products, Kampenhout, Belgium. This plasma was processed from the blood of healthy Sprague Dawley rats according to the certificate of the producer Equitech-Bio Inc. Citrated rat plasma "rat 2", prepared from pooled rat (mixture of strains) blood, lyophilized, product number P2516, Lot Nr. SLBN0606V, was obtained from Sigma-Aldrich GmbH, Steinheim. This plasma was reconstituted with 1 mL water as recommended by the provider. Plasma as well as spiked plasma samples were stored at -20°C.

*Other*

Aqueous solutions were prepared with HPLC grade water purchased from VWR, product No 23595.328. $MgCl_2$ product number 68475, Lot Nr. 1151053, was obtained from Sigma-Aldrich GmbH, Steinheim.

**Methods**

*Heparin Red® Kit*

Heparin, standard protocol (figure 1, 2; table 1)

For determination of heparin in plasma samples, the protocol of the provider for a 96-well microplate assay was followed with a minor modification: The mixture of Heparin Red solution and Enhancer solution was freshly prepared in 1:90 ratio (100 µL + 9 mL).

Heparin, protocol with added $MgCl_2$ (figure 3, 4; table 2)

As above, but the freshly prepared reagent mixture was 8.775 mL Enhancer solution + 225 µL aqeous $MgCl_2$ solution + 100 µL Heparin red solution.

Fucoidan, modifed protocol (figure 5, 6; table 3)

For determination of fucoidan in plasma samples, the protocol was modified as follows: The Heparin Red - Enhancer mixture was freshly prepared in 1:360 ratio (25 µL + 9 mL).

Fucoidan, modifed protocol with added $MgCl_2$ (figure 7, 8; table 1)

As above, but the freshly prepared reagent mixture was 8.775 mL Enhancer solution + 225 µL aqeous $MgCl_2$ solution + 25 µL Heparin Red solution.

All assays were continued as recommended by the protocol of the provider: 20 µL of the heparin or fucoidan spiked plasma sample was pipetted into a microplate well, followed by 80 µl of the Heparin Red – Enhancer ($MgCl_2$) mixture. For sample numbers > 10, a 12-channel pipette was used for addititon of the Heparin Red – Enhancer ($MgCl_2$) mixture. The microplate was introduced in the fluorescence reader and shaken for 3 minutes, using plate shaking function of the reader (setting "high"), followed by fluorescence measurement within 1 minute.

*Preparation of spiked plasma samples*

Plasma samples containing defined concentrations of heparins were prepared as follows:
Aqueous solutions (2 vol%) of unfractionated heparin and fucoidan, respectively, were added too pooled human plasma to achieve a concentration of 10 µg/mL. Concentrations required for the detections were adjusted by further dilution of this 10 µg/mL stock solution with the same plasma. The spiked plasma samples were stored at -20°C and thawed at room temperature before use.

*Data analysis*

Data were analyzed using Excel (Microsoft Office 10). Linear regression "through origin" ( with y-intercept set to 1) was applied to the response curves in figure 2, 4, 6 and 8. Coefficients of determination ($r^2$) in table 1-4 were determined by linear regression.

## Results and discussion

### Quantification of unfractionated heparin in mouse, rat and human plasma

In a first series of measurements, the Heparin Red Kit was applied to the determination of heparin in spiked plasma matrices (mouse, rat1, rat2 and human) by following the standard protocol recommended by the provider. The fluorescence response of Heparin Red is shown in figure 1. Obviously, the fluorescence of Heparin Red in the heparin-free rodent plasmas is much lower compared with human plasma. This could be related to association of Heparin Red with a plasma component of the rodent plasmas that is not prevalent in human plasma, leading to fluorescence quenching. While the response curves for mouse and rat1 plasma are almost overlapping, the difference between the plasmas rat1 (not lyophilized; Sprague Dawley rats) and rat2 (lyophilized and reconstituted; mixture of strains) is quite significant.

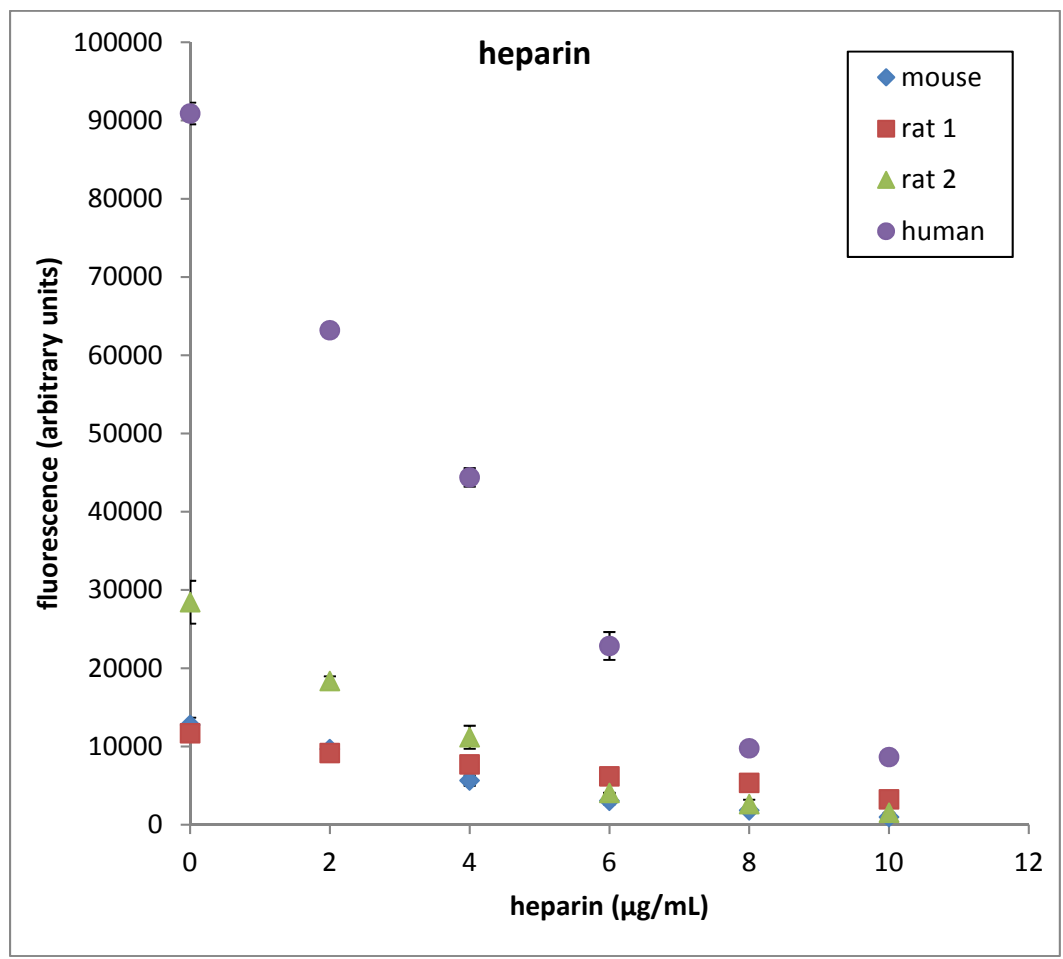

**Figure 1**. Fluorescence response (605 nm emission) of Heparin Red to heparin in four different plasma matrices: mouse, rat1, rat2 and human, using the protocol recommended by the provider (20 µl plasma sample is mixed with 80 µL reagent solution in a microplate well). Heparin in µg/mL refers to the concentration in spiked plasma samples. Manually performed microplate assays, as described in the "Materials and Methods" section. Averages of duplicate determinations.

In spite of very different fluorescence intensity of Heparin Red in the plasma matrices (figure 1), the *normalized* response to heparin (figure 2) is quite similar. Actually, response in the rodent plasmas is even better compared with human plasma. We speculate that fluorescence quenching of Heparin Red in the rodent plasmas is triggered by a weak, reversible interaction, and the binding equilibrium is largely shifted toward the strong complex with heparin (compare scheme 2). Moreover, human plasma proteins may bind heparin stronger [31] than rodent plasma proteins and weaken the response of Heparin Red to heparin in human plasma.

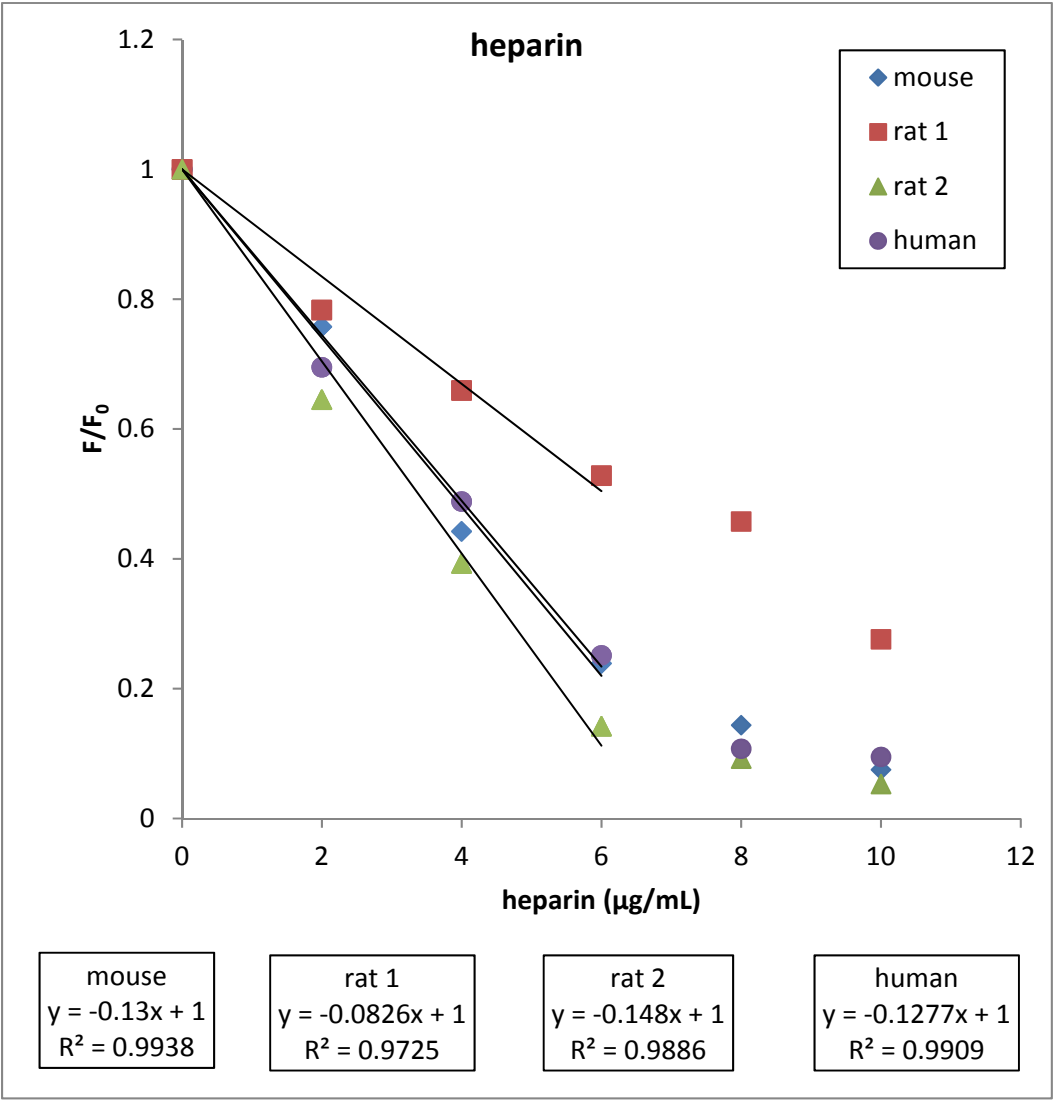

**Figure 2**. Data from figure 1, displayed as normalized fluorescence response ($F/F_0$) of Heparin Red to heparin in the four different plasma matrices: human, mouse, rat1 and rat2. Linear regression "through origin (heparin = 0 µg/mL; $F/F_0$ = 1)" was applied to the heparin concentration range 0-6 µg/mL.

The detection and quantification limits for the different plasma matrices (table 1) were determined based on signal-to-noise [32], by relating extrapolated response (linear regression "through origin" for the concentration range 0-6 µg/mL) to the standard deviation of blank samples ($\sigma_{blank}$) without heparin. The limit of detection (LOD) was calculated as LOD = 3 $\sigma_{blank}$ / S (S= initial slope of response curve, see figure 2) and the limit of quantification as LOQ = 10 $\sigma_{blank}$ / S.

| Plasma | Human | Mouse | Rat1 | Rat2 |
| --- | --- | --- | --- | --- |
| $\sigma_{blank}$ (n=8) | 0.014 | 0.117 | 0.112 | 0.096 |
| $r^2$ | 0.98 | 0.99 | 0.97 | 0.99 |
| LOD (µg/mL) | 0.33 | 2.71 | 4.06 | 1.95 |
| LOQ (µg/mL) | 1.11 | 9.04 | 13.54 | 6.50 |

**Table 1**. $\sigma_{blank}$, coefficient of determination ($r^2$), limit of detection (LOD) and limit of quantification (LOQ) for heparin in four different plasma matrices, relating to the data in figure 2. $\sigma_{blank}$ is the standard deviation of the normalized optical signal generated by the heparin-free plasma samples. $r^2$ is the coefficient of determination obtained from linear regression (figure 2). LOD = 3 $\sigma_{blank}$ / S (S= slope of response curve, see figure 2). LOQ = 10 $\sigma_{blank}$ / S.

The low fluorescence intensity of Heparin Red in the rodent plasmas (figure 1) and poor limit of detection and quantification due to high standard deviation of blank samples (table 1) prompted us to modify the standard detection protocol by varying the composition of the reagent solution that is mixed with the plasma sample. A significant improvement was achieved by the presence of 50 mM $MgCl_2$ in the reagent solution. Fluorescence of Heparin Red in the rodent plasmas increases significantly and nearly approaches the intensity observed in human plasma (figure 3). The effect is attributed to the $Mg^{2+}$ ions since fluorescence is not in the same way restored by 100 mM NaCl (same concentration of chloride ions). Possibly, the $Mg^{2+}$ cations mask anionic sites of the rodent plasma component that interact with Heparin Red and trigger the fluorescence decrease.

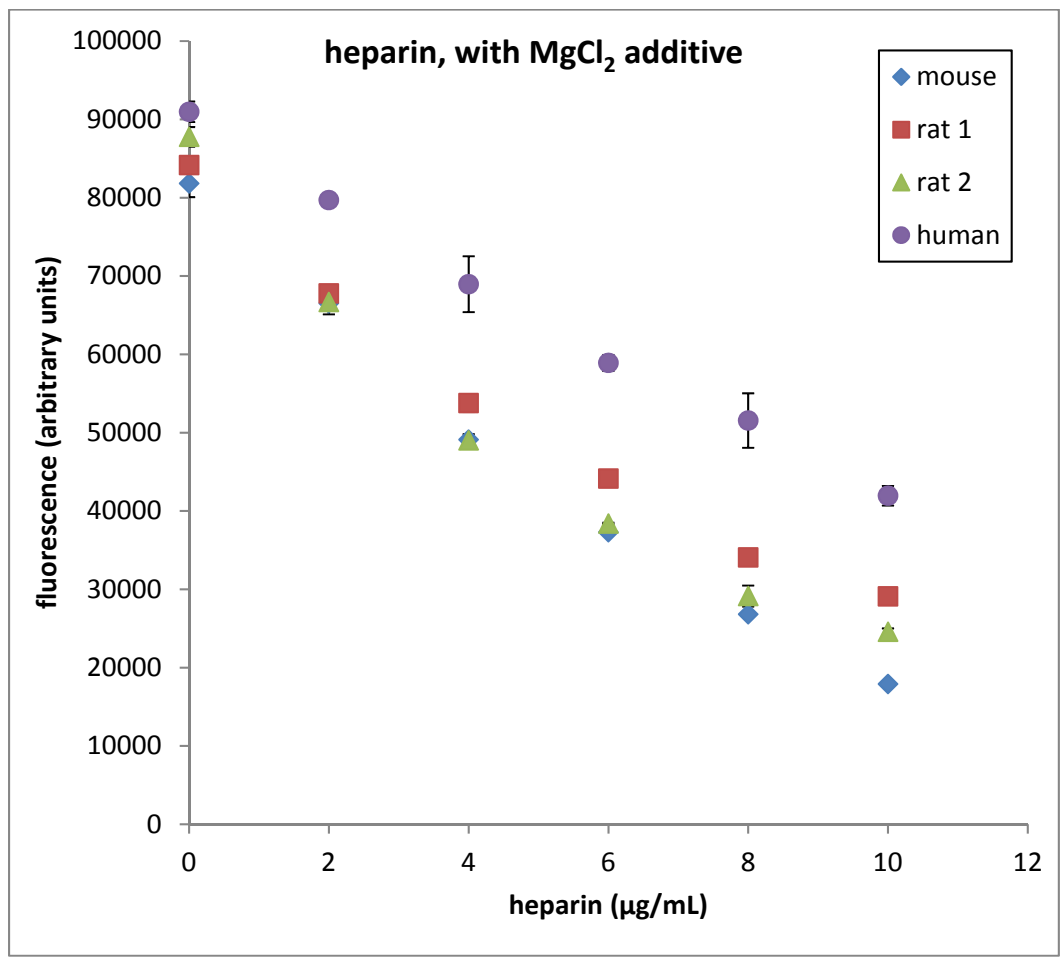

**Figure 3**. Fluorescence response (605 nm emission) of Heparin Red to heparin in four different plasma matrices: mouse, rat1, rat2 and human, using a modified protocol with a reagent solution containing 50 mM $MgCl_2$ (20 µl plasma sample is mixed with 80 µL reagent solution in a microplate well). Heparin in µg/mL refers to the concentration in spiked plasma samples. Manually performed microplate assays, as described in the "Materials and Methods" section. Averages of duplicate determinations.

Comparison of the normalized fluorescence response in the presence (figure 4) and absence (figure 2) of $MgCl_2$ indicates that the response of Heparin Red is somewhat diminished by $MgCl_2$. Divalent magnesium ions at concentrations above the physiological level were shown to compete with high-affinity proteins for heparin binding.[33] Similarly, high $Mg^{2+}$ concentrations present in the assay mixture may partially mask the interaction of Heparin Red with heparin.

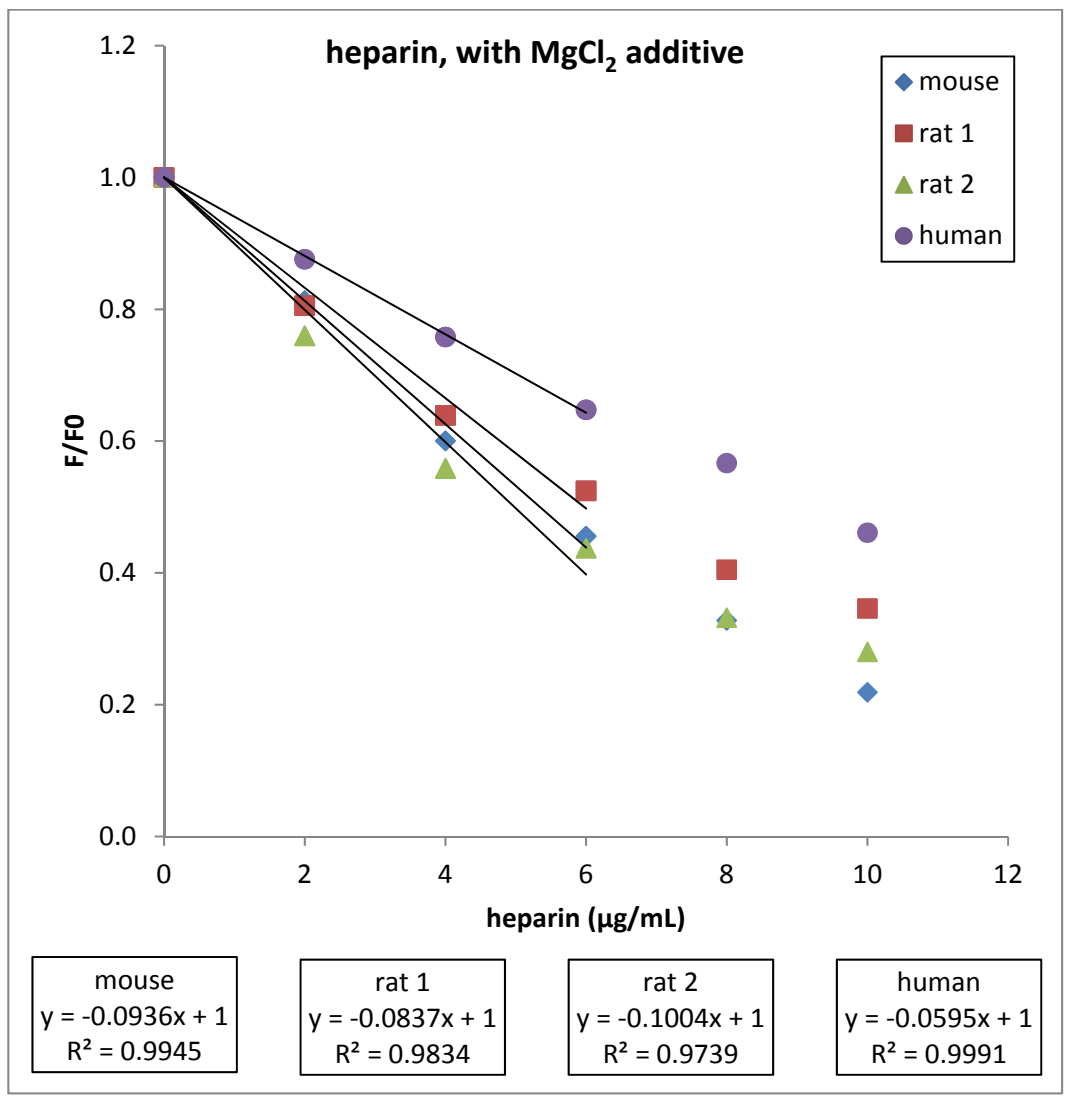

**Figure 4**. Data from figure 3, displayed as normalized fluorescence response (F/F$_0$) of Heparin Red to heparin in the four different plasma matrices: human, mouse, rat1 and rat2. Linear regression "through origin (heparin = 0 µg/mL; F/F$_0$ = 1)" was applied to the heparin concentration range 0-6 µg/mL.

A major beneficial effect of the MgCl$_2$ additive is the significant reduction of the standard deviation of fluorescence in the rodent plasma blank samples in the absence of heparin. Consequently, plasma heparin LOD and LOQ, since proportional to σ$_{blank}$, are very much improved (table 2), in spite of the diminished response.

| Plasma | Human | Mouse | Rat1 | Rat2 |
|---|---|---|---|---|
| $\sigma_{blank}$ (n=8) | 0.013 | 0.021 | 0.009 | 0.014 |
| $r^2$ | 0.999 | 0.99 | 0.98 | 0.97 |
| LOD (IU/mL) | 0.63 | 0.68 | 0.34 | 0.43 |
| LOQ (IU/mL) | 2.10 | 2.28 | 1.12 | 1.42 |

**Table 2**. $\sigma_{blank}$, coefficient of determination ($r^2$), limit of detection (LOD) and limit of quantification (LOQ) of heparin four different plasma matrices, relating to the data in figure 4 (modified protocol). $\sigma_{blank}$ is the standard deviation of the normalized optical signal generated by the heparin-free plasma samples. $r^2$ is the coefficient of determination obtained from linear regression (figure 4). LOD = 3 $\sigma_{blank}$ / S (S= slope of response curve, see figure 4). LOQ = 10 $\sigma_{blank}$ / S.

**Quantification of fucoidan in mouse, rat und human plasma**

Fucoidan from the brown alga *Fucus vesiculosus* has a lower averaged charge density per monosaccharide than heparin, -0.8 vs -1.8 (scheme 3). The weaker binding of Heparin Red to less highly charged polysaccharides leads to a diminished response relative to heparin in the competitive human plasma matrix.[20, 21, 34] The reduced sensitivity toward fucoidan in human plasma can be counteracted by a minor modification of the standard protocol, so that a similar concentration range as for heparin can be addressed. This modified protocol [21] uses a lower Heparin Red concentration (see "Materials and Methods" for details) and has also been applied to fucoidan detection in the present study. Otherwise, fucoidan determination in the different plasma matrices was performed and evaluated as described above for heparin.

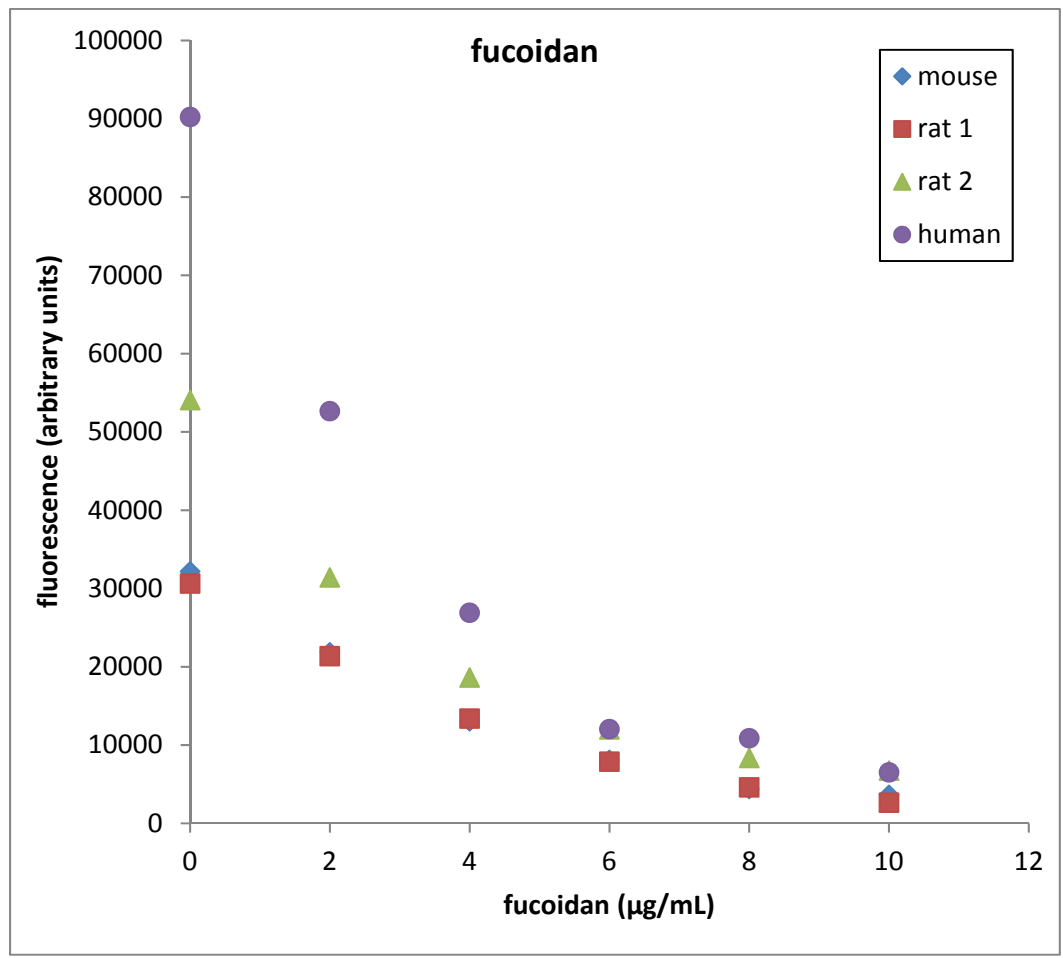

**Figure 5**. Fluorescence response (605 nm emission) of Heparin Red to *F. vesiculosus* fucoidan in four different plasma matrices: mouse, rat1, rat2 and human, using a modified protocol (see "Materials and Methods" for details; 20 µl plasma sample is mixed with 80 µL reagent solution in a microplate well). Fucoidan in µg/mL refers to the concentration in spiked plasma samples. Manually performed microplate assays, as described in the "Materials and Methods" section. Averages of duplicate determinations. Sensitivity of the microplate reader has been adjusted (gain 120).

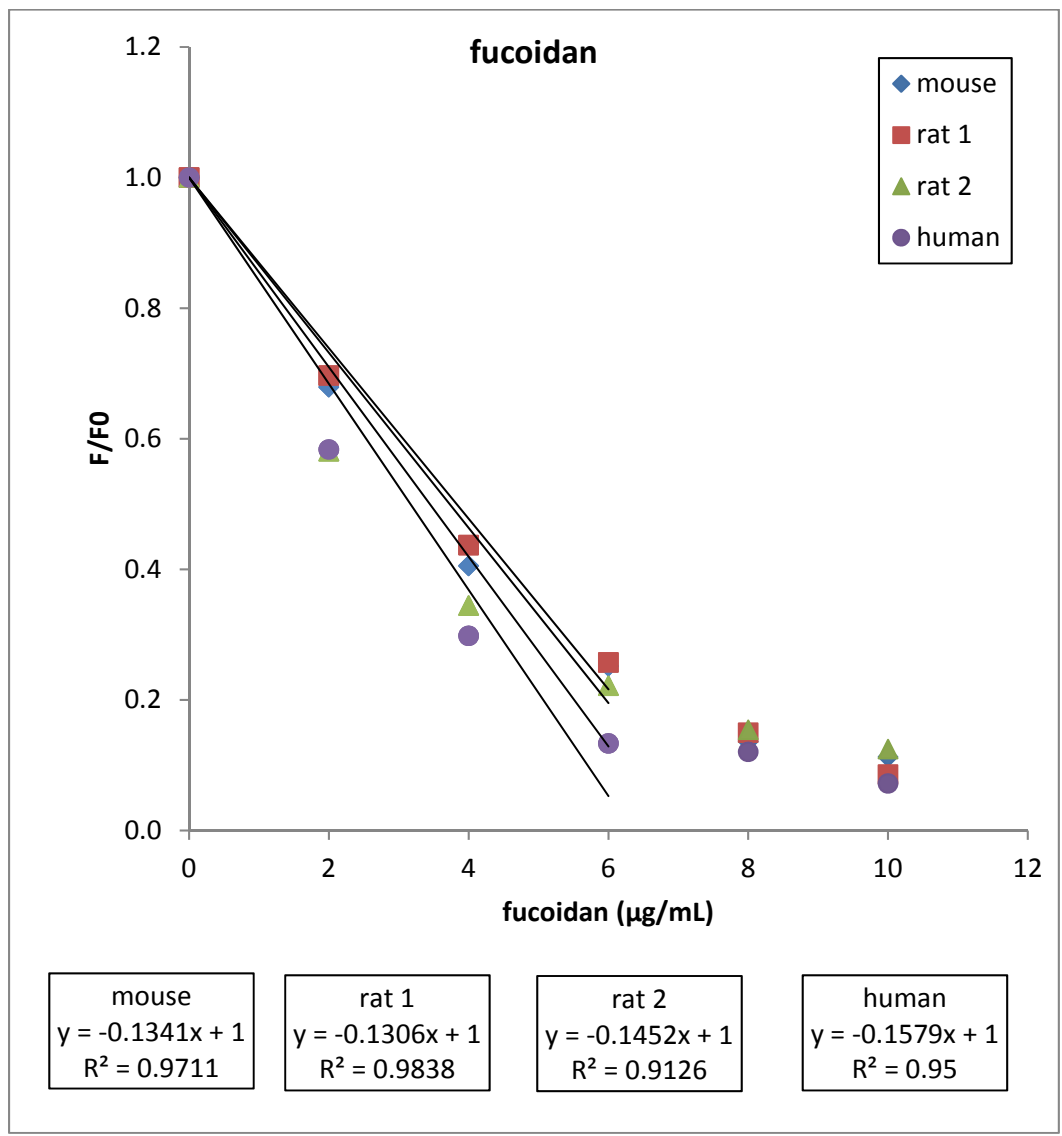

**Figure 6**. Data from figure 5, displayed as normalized fluorescence response (F/F$_0$) of Heparin Red to fucoidan in the four different plasma matrices: human, mouse, rat1 and rat2. Linear regression "through origin (heparin = 0 µg/mL; F/F$_0$ = 1)" was applied to the heparin concentration range 0-6 µg/mL.

| Plasma | Human | Mouse | Rat1 | Rat2 |
|---|---|---|---|---|
| σ$_{blank}$ (n=8) | 0.031 | 0.046 | 0.071 | 0.03 |
| r$^2$ | 0.95 | 0.97 | 0.98 | 0.91 |
| LOD (IU/mL) | 0.60 | 1.03 | 1.63 | 0.62 |
| LOQ (IU/mL) | 1.97 | 3.45 | 5.42 | 2.06 |

**Table 3**. σ$_{blank}$, coefficient of determination (r$^2$), limit of detection (LOD) and limit of quantification (LOQ) for fucoidan in four different plasma matrices, relating to the data in figure 6. σ$_{blank}$ is the standard deviation of the normalized optical signal generated by the heparin-free plasma samples. r$^2$ is the coefficient of determination obtained from linear regression (figure 6). LOD = 3 σ$_{blank}$ / S (S= slope of response curve, see figure 6). LOQ = 10 σ$_{blank}$ / S.

Fluorescence response follows similar trends as for heparin: Heparin Red fluorescence is lower in the rodent plasmas (figure 5), but normalized response to fucoidan comparable for all four plasmas (figure 6). Compared with the standard protocol for heparin (figure 2, table 1), standard deviation of the rodent plasma blanks is significantly smaller (table 3) for this modified protocol using a lower Heparin Red concentration, resulting in lower LOD and LOQ. Addition of $MgCl_2$ again leads to a siginificant enhancement of Heparin Red fluorescence in the rodent plasmas (figure 7), along with a somewhat diminished response. The beneficial effect of $MgCl_2$ on $\sigma_{blank}$, LOD and LOQ (table 4 vs table 3) for fucoidan detection is present but less pronounced than for heparin detection (table 2 vs table 1).

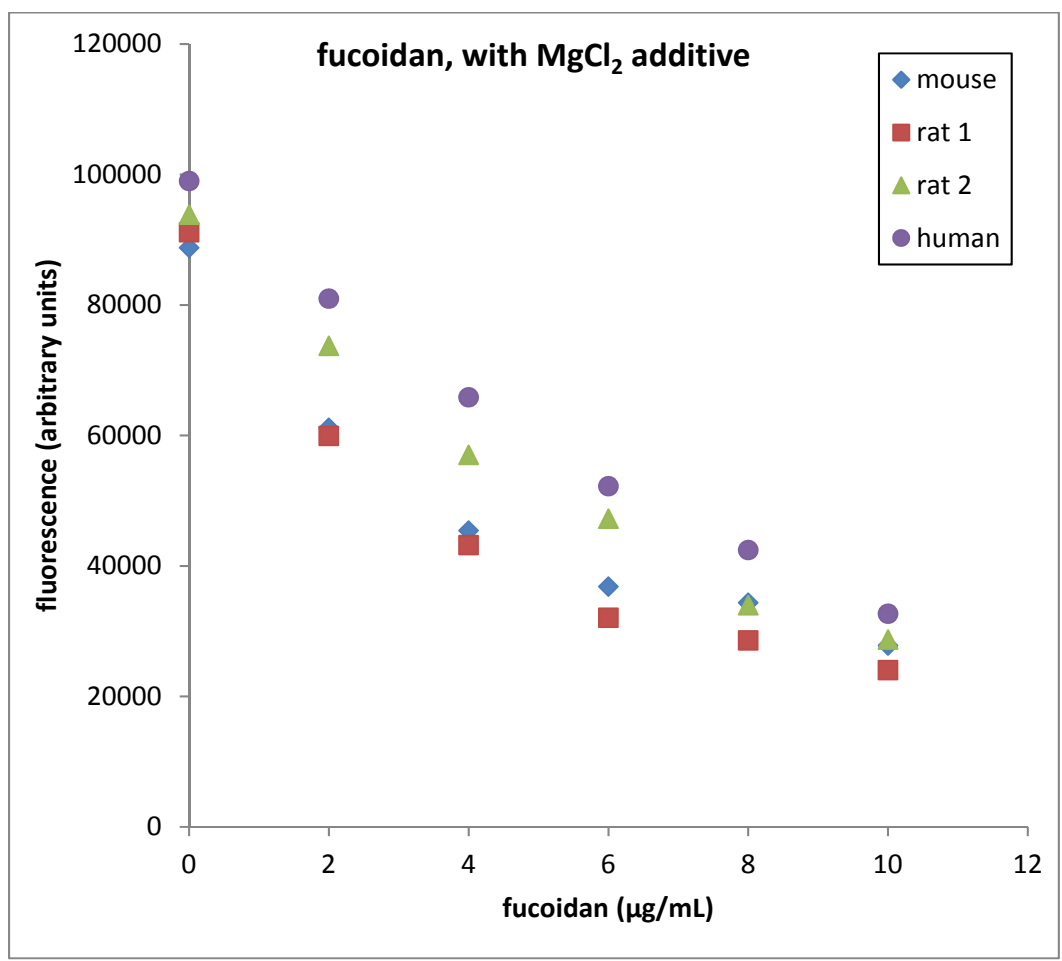

**Figure 7**. Fluorescence response (605 nm emission) of Heparin Red to heparin in four different plasma matrices: mouse, rat1, rat2 and human, using a modified protocol with a reagent solution containing 50 mM $MgCl_2$ (see "Materials and Methods" for details; 20 µl plasma sample is mixed with 80 µL reagent solution in a microplate well). Fucoidan in µg/mL refers to the concentration in spiked plasma samples. Manually performed microplate assays, as described in the "Materials and Methods" section. Averages of duplicate determinations. Sensitivity of the microplate reader has been adjusted (gain 120).

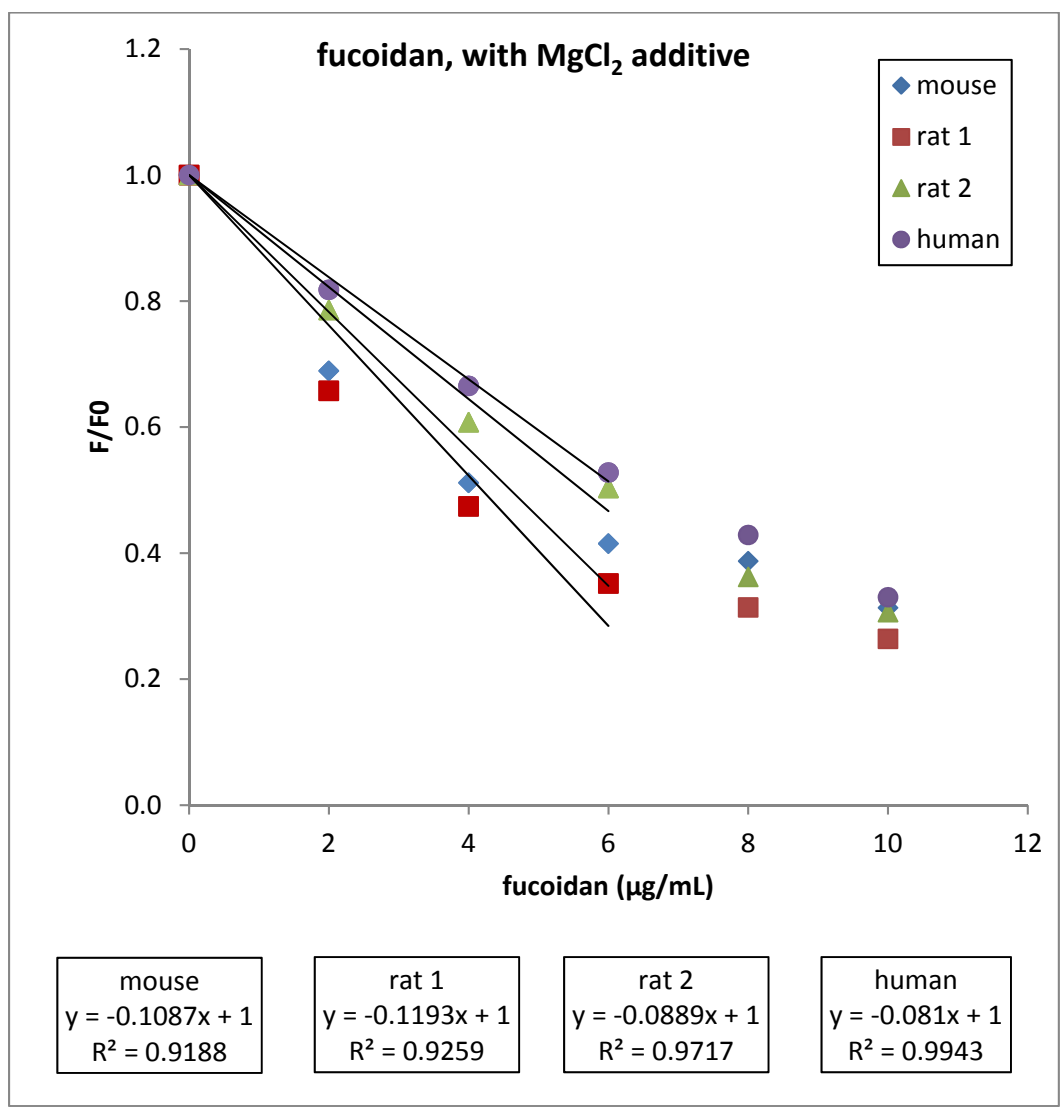

**Figure 8**. Data from figure 7, displayed as normalized fluorescence response (F/F$_0$) of Heparin Red to fucoidan in the four different plasma matrices: human, mouse, rat1 and rat2. Linear regression "through origin (heparin = 0 μg/mL; F/F$_0$ = 1)" was applied to the heparin concentration range 0-6 μg/mL.

| Plasma | Human | Mouse | Rat1 | Rat2 |
|---|---|---|---|---|
| σ$_{blank}$ (n=8) | 0.015 | 0.037 | 0.023 | 0.015 |
| r$^2$ | 0.99 | 0.92 | 0.92 | 0.97 |
| LOD (IU/mL) | 0.55 | 1.01 | 0.58 | 0.51 |
| LOQ (IU/mL) | 1.84 | 3.36 | 1.92 | 1.70 |

**Table 4**. σ$_{blank}$, coefficient of determination (r$^2$), limit of detection (LOD) and limit of quantification (LOQ) for fucoidan in four different plasma matrices, relating to the data in figure 8. σ$_{blank}$ is the standard deviation of the normalized optical signal generated by the fucoidan-free plasma samples. r$^2$ is the coefficient of determination obtained from linear regression (figure 8). LOD = 3 σ$_{blank}$ / S (S= slope of response curve, see figure 8). LOQ = 10 σ$_{blank}$ / S.

## Conclusion

This study addresses the need for simple and user-friendly analytical methods for quantifying the blood levels of sulfated polysaccharides in mice and rats in the context of preclinical drug development. The commercially available, direct fluorescence assay Heparin Red was applied to the quantification of heparin, a highly sulfated polysaccharide, and fucoidan, a less sulfated polysaccharide, in spiked mouse and rat blood plasma samples. Application of a standard protocol for human plasma resulted in poorer detection sensitivities for the rodent plasma matrices. Quantification limits (LOQs) for both target analytes could be improved significantly by simple modifications of the protocol, in particular by the addition of magnesium chloride to the assay mixture. The improved protocol provides within the detection range 0-10 µg/mL the LOQs 1.1 – 2.3 µg/mL for heparin, and 1.7 – 3.4 µg/mL for fucoidan (LOQ is defined as 10 $\sigma_{blank}$ / S, $\sigma_{blank}$ being standard deviation of blank and S the initial slope of response curve). The low plasma sample volume of only 20 µL is advantegous in particular when blood samples are collected from mice.

Heparin Red is a promising tool for the preclinical evaluation of sulfated polysaccharides that hold promise as new therapeutics. Polysaccharides of varying sulfation degrees are determined at low µg/mL levels in mouse and rat plasma by a simple mix-and-read microplate assay.

**Conflict of interest.** R. Krämer holds shares in Redprobes UG, Münster, Germany. Other authors: No conflict of interest.